# Criteria for realizing room temperature electrical transport applications of topological materials

*Matthew Brahlek*[1]*

[1]Materials Science and Technology Division, Oak Ridge National Laboratory, Oak Ridge, TN, 37831, USA

Email: *brahlekm@ornl.gov

**Abstract**: The unusual electronic states found in topological materials can enable a new generation of devices and technologies, yet a long-standing challenge has been finding materials without deleterious parallel bulk conduction. This can arise either from defects or thermally activated carriers. Here, I clarify the criteria that materials need to meet to realize transport properties dominated by the topological states, a necessity for a topological device. This is demonstrated for 3-dimensional topological insulators, 3D Dirac materials, and 1D quantum anomalous Hall insulators, though this can be applied to similar systems. The key parameters are electronic band gap, dielectric constant, and carrier effective mass, which dictate under what circumstances (defect density, temperature, etc.) the unwanted bulk state will conduct in parallel to the topological states. As these are fundamentally determined by the basic atomic properties, simple chemical arguments can be used to navigate the phase space to ultimately find improved materials. This will enable rapid identification of new systems with improved properties, which is crucial to design new materials systems and push into a new generation of topological technologies.

Topology is a very general notion used to broadly classify objects, and has been extensively used in condensed matter physics to describe many systems including, the relevant system of the current work, topological band structures[1,2]. Examples include the 1-dimensional (1D) states that exist on the boundary of a 2-dimensional (2D) insulator such as the quantum Hall insulator (QHI)[3,4], quantum spin Hall phase[5,6], and quantum anomalous Hall insulator (QAHI)[7–10], the 2D Dirac-like topological surface states (TSS) that exist on the surfaces of 3D topological insulators (TIs)[11] or topological crystalline insulators (TCIs), and 3D Dirac systems such as 3D Dirac semimetals (DSM) or Weyl semimetals. Novel linearly dispersing bands characteristic of these topological states are highlighted in red in Figure 1(a-c). Together the unusual character of topological materials make them of interest for a wide range of fundamental studies and many applications in areas ranging from quantum computation to spintronics[12–14]. However, one of the major challenges is that the experimentally realizable materials have defects and are rarely true insulators—their trivial 2D or 3D bulk states, highlighted in Figure 1(a-c) as blue (2D) and black (3D) lines, are nearly ubiquitously metallic, which acts as a parallel conduction channel that electrically shorts the transport through the topological states.

The origin of parallel bulk conduction coincides with several aspects that drive materials into a topological phase, which impacts the conditions where the topological states may dominate the transport. For a 3D TI, the $Z_2$ topological number becomes non-trivial only when the bulk conduction and valence bands are inverted due to strong spin-orbit coupling. This requires narrow band gaps where spin-orbit coupling is sufficiently strong to invert the orbital character of the conduction and valence bands (spin-orbit coupling, in a broad sense is relatively weak compared the kinetic energies and Coulombic interactions responsible for bonding and crystallization, and, thus, band gap formation). This requirement is fulfilled when the constituent elements have similar electronegativity, giving rise to a strong covalency[15]. This means that the band gap will be narrow and charged antisite defects will have a low formation energy, and, therefore, are likely to occur with a high density. The key materials that have dominated the field over the past decade are the tetradymite 3D TIs $Bi_2Se_3$, $Bi_2Te_3$, and $Sb_2Te_3$[15,16], and it has proven a major challenge to both understand and mitigate defects sufficiently to achieve a true topological insulating phase from transport[17–21]. However, the caveat arises that this true insulating phase for topological materials occurs at very low temperatures near absolute zero, often requiring dilution refrigerators operating at sub-Kelvin temperatures. This highlights an additional problem resulting from the intrinsically narrow bandgap: a substantial density of thermally activated bulk carriers arises even at low temperatures. This is a challenge for many researchers to probe these states as well as precluding or hampering any technological applications. Here, I show that the band gap, dielectric constant, and effective mass combine to dictate under what conditions the topological states may dominate the electronic response of a material. This is determined by the critical carrier density when the bulk states become metallic and have conductivity sufficiently high to obscure the transport through the topological states. Examples are 3D TIs and TCIs, 3D Dirac systems and the 1D QAHI. Beyond these, this is broadly applicable and provides well-defined bounds for where transport properties can be dominated by the topological states. These parameters are fundamentally tied to key aspects of the materials chemistry, which will serve as a guide to find new materials that will enable room temperature topological devices.

For a pristine undoped semiconductor the Fermi level is located within the band gap, and it is said to be in the intrinsic limit. When it is doped either intentionally or unintentionally, in the extrinsic limit, the Mott-criterion gives the critical density of charged defects to push a 3D insulator into a metal phase with the Fermi level occupying either the conduction or valence band. When a dopant atom is incorporated into an insulator, its associated charge is screened by polarizing the surrounding atoms in the lattice. This screening follows a hydrogen-like potential, which is rescaled due to the dielectric response of the material. This yields an effective Bohr radius given by $a = \epsilon(m_e/m^*)a_B$ where $m_e$ is the free electron mass, $m^*$ is the effective mass, $\epsilon$ is the dielectric constant, and $a_B$ is the free-space Bohr radius which is about 0.5 Å. The Mott-criterion predicts that when atoms are doped into a material with a sufficient density, $N_D$, then the insulator becomes metallic. This critical density is given by $N_D \approx (0.26/a)^3$ [22]. For the tetradymite TIs ($Bi_2Se_3$, $Bi_2Te_3$, and $Sb_2Te_3$) the dielectric constants are relatively large, ~100-300, and the effective masses are relatively small, ~$0.2m_e$[16]. This yields a relatively low $N_D \approx 10^{14}$-$10^{15}$ cm$^{-3}$ [16,19]. Further, this range of parameters that is typical for topological materials has been implicated in the poor screening of defects and compensation dopants, which gives rise to weakened insulating behavior with thermal activation energies much smaller than expected[23,24]. In contrast, for semiconductor materials like Si the critical density is orders-of-magnitudes larger at around $10^{17}$-$10^{18}$ cm$^{-3}$ [22], owing to the larger effective mass and smaller dielectric constant. Finite disorder, due to effects such as Anderson localization, however, aid the suppression of these bulk states. This is captured by the Ioffee-Regel criterion where $k_F l \sim 1$, where $k_F$ is the Fermi wave vector and $l$ is the mean-free-path, which will push the insulator to metal transition to higher density[19]; therefore, $N_D$ is taken to be $10^{15}$ cm$^{-3}$ for most TIs and TCIs (the Mott criteria is not applicable to DSM and Weyl semimetals since the Dirac states are 3D and metallic).

A topological device must operate at finite temperature, which is ideally room temperature. To establish an upper limit on the temperature scale for transport to be dominated by a topological state requires assuming the material is in the intrinsic limit, for which bulk carriers must be induced thermally. This resulting bulk conductivity can then be compared to that of the topological states. In other words, the Fermi level, $E_F$, is located at the center of the band gap (i.e. $N_{3D} < N_D$), and finite temperature will then cause carriers to be excited across the bulk band gap. The number density of thermally induced carriers is then given by

(1) $$N(T) = \int D(E) f(E,T) dE,$$

where $D(E)$ is the density of states and

(2) $$f(E,T) = \frac{1}{1+e^{E/(k_B T)}}$$

is the Fermi function, where $E$ is the energy, $T$ the temperature and $k_B$ is Boltzmann's constant. The density of states is simply $D_{d,j}(E) = \frac{dn_d}{dE} = \frac{dn_d}{dk}\frac{dk}{dE_j}$, where $n_d$ is the number of states for a given wavevector, $k$, with dimensionality $d$, $E$ is the energy dispersion, and $j$ indexes the scaling of the dispersion, i.e. $E \propto k^j$ (where $j = 1$ for a Dirac system and $j = 2$ for a parabolic dispersion). For example, in 3D and for a quadratic dispersion $n_{3D} = \frac{g}{(2\pi)^3}\frac{4\pi}{3}k_F^3$ and $E = \frac{\hbar^2 k^2}{2m_*}$, which yields

(3) $$D_{3D,i}(E) = \frac{g}{4\pi^2}\left(\frac{2m_i^*}{\hbar^2}\right)^{3/2}\sqrt{E - \frac{E_g}{2}}.$$

Here, $g$ is the spin degeneracy ($g = 2$ for a spin degenerate band and 1 for a gapped TSS), $\hbar$ is Planck's reduced constant, $E_g$ is the band gap, and the subscript $i$ indicates if the band is an electron ($n$) or hole ($p$). In 3D and for a quadratic dispersion, the density of carriers is then calculated by evaluating

$$(4) \quad N_{3D}(T) = \int_{-\infty}^{-\frac{E_G}{2}} D_{3D,p}(E)f(E,T)dE + \int_{\frac{E_G}{2}}^{\infty} D_{3D,n}(E)f(E,T)dE.$$

As schematically shown in Figure 1, topological band structures are characterized by band gaps with a linearly dispersing state within this gap: For a 3D TI with a band gap $E_G$, the gap, formed by the conduction and valence bands (black lines), is spanned by 2D Dirac like TSS (red lines). If the 2D TSS can be gapped, $\Delta$, via breaking, for example, time reversal symmetry, this can give rise to a linear 1D chiral edge state characteristic of the QAHI (red solid line). As such, Equation (1) can be evaluated across a range of dimensions, dispersions, and realistic materials parameters to understand the bounds on temperature for observing transport effects dominated by these novel states. This gives the number of free carriers, and, although simple, can give a good estimation on the bounds where the topological states dominate the transport processes. As such, complexities were not considered but can easily be included in future materials-specific calculations. This includes differences in the valence band and conduction band masses, shifts of the band gap versus temperature [25], band bending effects[19], spatial fluctuations due to charged dopants[23,24], multiple band valleys or Dirac bands as well as linear corrections to the dispersion of the bulk bands[25]. First, the case of 3D materials is discussed, followed by the case of a gapped TSS and the associated QAHI.

Although closed solutions likely can be found for Equation (1) with some limiting assumptions (e.g. $E_G \gg k_BT$ reduces Fermi statistics to Boltzmann statistics), numerical solutions were deemed sufficient since the number and range of realistic parameters is small. The free parameters are the band gap and the effective mass. As mentioned above, the current host of topological materials are typically around $m^* \approx 0.2 m_e$ with band gaps less than about 0.3 eV (see Ref. [26] for a thorough list of known and predicted TIs). Equation (4) was evaluated numerically, and, as shown in Figure 2, the resulting $N_{3D}$ curves were plotted as a function of temperature for $m^* \approx 0.2 m_e$ with band gaps ranging from 0.01-1.1 eV. For consistency, it is noted that these results agree well with Ge ($E_G$ = 0.7 eV) and Si ($E_G$ = 1.1 eV) that have intrinsic carrier densities at room temperature of $N_{3D,Ge} \approx 2\times10^{13}$ cm$^{-3}$ and $N_{3D,Si} \approx 1\times10^{10}$ cm$^{-3}$[27], where $N_{3D}$ in Figure 2 is a slight underestimate considering the effective masses of Si and Ge are larger than $0.2m_e$, see Figure 3(b). From this calculation, the number of free carriers ubiquitously increases with increasing temperature with the lowest band gap having the steepest increase, as expected. For comparison, an extrinsic carrier density of the approximate Mott criterion of $N_D \approx 10^{15}$ cm$^{-3}$ is plotted as a dashed gray horizontal line. As such, if the defect density is at or below $N_D$, thermally activated carriers may be the dominant source of bulk conduction. The temperature at which this occurs will, however, depend on the band gap, which approximately occurs where the thermally activated carrier curves intersect $N_D$. This occurs for temperatures less than 150 K for $E_G$ < 0.3 eV, which is of order of the largest band gap for confirmed topological materials.

Designing or modifying materials to mitigate deleterious conduction from non-topological states necessitates understanding how the total conductivity varies with temperature, band gap, and effective mass, etc. This comparison can be made by calculating the percent of conductance that arises from the Dirac state relative to the total, $P_{DS-BS}(\%) = G_{DS}/(G_{DS} + G_{BS}) \times 100\%$. Here, $G_{DS}$ is the conductance of the Dirac state, either 2D, $G_{DS,2D} = e\mu_{DS}n_{2D,DS}$, or 3D, $G_{DS,3D} = e\mu_{DS}N_{3D,DS}t$, with thickness $t$, and $G_{BS} = e\mu_{BS}N_{3D}t$ is the conductance from the bulk state. This can be rearranged to yield $P_{DS-BS}(\%) = 1/\left(1 + \alpha \frac{N_{3D}t}{n_{2D,DS}}\right) \times 100\%$ for the 2D Dirac state and $P_{DS-BS}(\%) = 1/\left(1 + \alpha \frac{N_{3D}}{N_{3D,DS}}\right) \times 100\%$ for the case of a 3D Dirac state, where $\alpha$ is the ratio of the mobilities. Assuming $\alpha$ is of order unity (this is empirically valid for the tetradymite TIs where bulk and TSS are typically both of order of several thousand cm$^2$V$^{-1}$s$^{-1}$), $P_{DS-BS}$ can be calculated to elucidate how the maximum conductivity arises through the topological states. The results for the temperature fixed at $T$=300 K are examined. To evaluate $P_{DS-BS}$, it is again assumed that $E_F$ is near the Dirac point at the center of the band gap, and, hence, $N_{3D}$ is given by the results in Figure 2. The number of carriers originating from the Dirac band can be similarly calculated using Equation (1) with the density of states calculated with linear dispersion for 2D and 3D. The only parameter is then the degeneracy $g$ and the Fermi velocity, which is of order $5\times10^5$ ms$^{-1}$ for most topological materials. This yields $n_{2D,DS} \approx 1\times10^{11}$ cm$^{-2}$ with $g = 1$ for the tetradymite TIs, and $N_{3D,DS} \approx 1\times10^{17}$ cm$^{-3}$ with $g = 2$ for a DSM such as Cd$_3$As$_2$. The calculations of $P_{DS\text{-}BS}$ for the 2D case were performed with $t = 20$ nm. This value was selected since it is the thickness where the TSS are decoupled for a TI with band gap of 0.3 eV in transport measurements[18,28]; and this thickness is roughly equivalent to a 3D Dirac state with $N_{3D,DS} \approx 1\times10^{17}$ cm$^{-3}$, since the conductance is approximately the same.

The results of these calculations are shown in the lower panel of Figure 3(a), where $P_{DS\text{-}BS}$ is plotted versus $E_G$ for various effective masses. For comparison, horizontal dashed lines indicate the tolerance to bulk conduction at $P_{DS\text{-}BS} = 95\%$ and 99%. For further comparison, the top panel of Figure 3(a) shows select data for various topological materials including TIs, TCIs, and DSMs. Based on the band gaps and approximate effective masses of these materials, this plot shows approximately less than half of the conduction will be through the topological states at room temperature. For the case of a 3D TI, reducing thickness from 20 nm to 5 nm, the minimum value where surface states hybridize[29], will only produce marginal improvements. This explicitly shows that to realize room temperature transport dominated by the topological states requires modifying the materials parameters. This calculation, in fact, gives guidance to navigate the space of parameters where these ideal properties might be realized. Taking $P_{DS\text{-}BS} = 95\%$, 99%, and 99.9% enables finding the set of band gaps and effective masses where the conductance is dominated by the topological states. The results are plotted in Figure 3(b) where the horizontal axis is the band gap, and the vertical axis is the effective mass and the respective curves represent $P_{DS\text{-}BS} = 95\%$ (blue), 99% (green), and 99.9% (red); $P_{DS\text{-}BS} = 99.9\%$ was also included with $T = 77$ K (dashed red; here $n_{2D,DS} = 1\times10^{10}$ cm$^{-2}$). For a given band gap, the shaded area below these curves yields the set of parameters for which the conductance is dominated by the topological states. For example, given the band gaps of current topological materials $E_G \lesssim 0.3$ eV, ideal properties can be achieved if effective mass is reduced substantially—roughly an order of magnitude. Alternatively, for the current effective masses $m^* \gtrsim 0.1 m_e$, a band gap larger than around 0.5 eV would be required. For comparison, however, many current topological materials (e.g. $Sb_2Te_3$ and $Bi_2Se_3$) are well suited for the temperature of 77 K as indicated by the dashed red curve in Figure 3(b).

The QAHI is considered by numerically evaluating Equation (1), where the 3D band gap is replaced by the 2D magnetic gap $\Delta$, and the spin degeneracy term is taken to be $g = 1$, reflecting the spin polarized TSS. The results are plotted for effective masses of the 2D gapped band equal to $0.1m_e$ and $0.2m_e$ in Figure 4(a-b), respectively for various magnetic gaps ranging from 0.01-0.4 eV. One challenge arises for the case of 2D in comparison to the previous case of 3D: scaling theory predicts that all 2D systems should be localized[30]. This precludes the establishment of an analogy to the Mott-criterion. For comparison to the quantized resistance expected for the QAHI, a range of carrier densities were considered and the corresponding resistance relative to the quantized value was estimated. A mobility of order 2000 $cm^2V^{-1}s^{-1}$, which is typical of TSS[17,18,20,31] with carrier densities around $N_{2D} = 10^8$ $cm^{-2}$, $10^9$ $cm^{-2}$, and $10^{10}$ $cm^{-2}$ correspond to roughly $1000 \times h/e^2$, $100 \times h/e^2$ and $10 \times h/e^2$, respectively, and, therefore, perfect quantization would be obscured by about 0.1%, 1% and 10%. Similarly, $10^{11}$ $cm^{-2}$ corresponds to roughly $1 \times h/e^2$, which, for practical purposes, would wash out any signature of quantized transport. Contrary to the expectation that higher mobility is better, samples with larger mobility would have a lower resistance, and, therefore, the quantum transport would be more obscured, thus lowering the threshold. Plotted in Figure 4(c-d) are the temperatures where $N_{2D} = 10^8$ $cm^{-2}$ vs $\Delta$ for this lower limit as well as larger values. Realizing quantized transport at room temperature would require a magnetic gap in excess of 0.3 eV, which in terms of magnetic transitions corresponds to Curie temperatures well over 1000 K. Such temperatures are a challenge for intrinsic systems, yet may be possible for heterostructures with ferromagnetic insulators such as $Y_3Fe_5O_{12}$ (YIG), which has a magnetic transition around 560 K. However, for the current TIs the magnetic gap would have to be the same as the bulk band gap, making room temperature realization an ambitious goal. Nevertheless, it is well within the bounds of liquid nitrogen temperatures, especially considering heterostructures with high-temperature ferromagnets where gaps have been calculated in the range of 0.04-0.1eV[32,33].

The results obtained here give realistic bounds for where topological states may dominate the electrical response of a material. This depends on simple material parameters that are related to basic aspects of the materials chemistry. Understanding this dependence is the key to finding systems that are more robust to deleterious effects of defects and may operate at elevated temperatures that could be as high as room temperature.

Concerning 2D and 3D Dirac systems. For the topological states to dominate the transport requires finding materials with (1) *larger band gaps,* (2) *smaller dielectric constants*, and (3) *lower effective mass* (in the intrinsic limit, though higher effective mass improves robustness to defects, see below). Finding new materials with such improved properties is, however, a challenge since these parameters are strongly tied to the topological properties. Therefore, a careful balance is needed for each parameter individually.

(1) *Larger band gaps*: To find materials with larger band gaps requires a larger difference in electronegativity and increasing the ionic nature of the bonds[15]. This can be achieved, for example, by choosing anions with larger electronegativity, e.g. in the tetradymites replacing Te and Se with S, which should lower the valence band energy. This is demonstrated for $Bi_2Te_3$ and $Sb_2Te_3$, whose band gaps are narrower than $Bi_2Se_3$ and the ternary and quaternary tetradymites involving S. Doing so, however, simultaneously pushes the materials towards a topologically trivial phase. This is shown in Figure 3(a), where, once band inversion is destabilized, the band gap rapidly opens. This is explicitly clear for the non-TI $Bi_2Se_2S$, which has a band gap >1 eV. This implies that $Bi_2Se_3$ is right on the edge of being non-topological, yet the band gap is only 0.3 eV. This poses the question: is there a fundamental maximum for band gaps of topological materials, and can the necessary band gap of 0.5 eV be realized?

(2) *Smaller dielectric constants*: Reducing the dielectric constant would increase the critical density at which the bulk states would become metallic, and, thus, is critical in reaching the intrinsic limit. Reducing the dielectric constant by a factor of two or better, to below 50, will boost $N_D$ by nearly an order of magnitude, which dramatically improves robustness to the effect of defects. Achieving such improvements can be accomplished by considering the factors that determine the dielectric response of a material[34]. (a) Electronic polarizability due to the displacement of the electron cloud relative to the nuclei:

Atoms with larger electron density have higher polarizabilities, so materials with heavy elements like Bi tend to have high dielectric constants. This can be seen in the tetradymite TIs where $Bi_2Te_3$ has a dielectric constant that is nearly twice that of $Bi_2Se_3$[16]. This highlights that materials that are likely topological will likely have large dielectric constants, and, thus, low $N_D$. (b) Ionic polarizability results from shifts in the cation and anion positions due to their net charge, and, therefore, the more ionic the bond the higher the dielectric constant. This is, however, not likely to be a major concern since compounds that favor the narrow band gaps necessary for band inversion are typically highly covalent (ionic materials tend to have larger gaps). More covalent character is typically favored in the p-block of the periodic table; as discussed below, this simultaneously favors lower effective masses. (c) Dipole polarizability occurs when a preexisting dipole changes its orientation in the presence of an electric field. In solids, these are either ferroelectrics or in close proximity to ferroelectricity, and are typically rare, and should be avoided for TIs in the current context[35].

This overall trend is captured in Figure 5, which shows a compilation of dielectric constants versus band gap across a wide range of materials including both topological materials and non-TIs. It is noted that extracting the zero-frequency measurements of dielectric constant for materials whose bulk is not strongly insulating is a challenge, which involves extrapolating to zero frequency using an oscillator model[36]. This is the case for most topological materials. Generally, as the net mass of the elements is lowered, the dielectric constant is also lowered. There is also a strong correlation with band gap, which tends to increase as the mass decreases. The important trend among band gap and dielectric constant is captured empirically as $\epsilon = 25E_G^{-0.6}$[37]. Open questions remain regarding how large of a band gap and how small of a dielectric constant can be achieved for a topological material. However, this relation gives optimism, because a larger band gap should simultaneously imply lower dielectric constant. Given that materials with masses similar to topological materials, such as GaSb and InSb, can reach $\epsilon < 50$, it is likely that topological materials can be found with significantly reduced dielectric constants. Achieving this will impart a robustness to defects and enable access to the intrinsic limit where the topological states may dominate the transport.

(3) *Lower effective mass* (in the intrinsic limit): The dependence on effective mass is more complex as it is involved in both the density of thermally activated carriers and the Mott criterion, both of which have opposite behaviors. In the intrinsic limit, lowering the effective mass raises the contribution to the conductance from the topological states. This can be seen by considering Equation (4-5) where $N_{3D} \propto m^{*3/2}$. As such, $P_{DS-BS} \propto 1/(1+\gamma m^{*3/2})$, where $\gamma$ depends on band gap, temperature, etc. Therefore, $P_{DS-BS}$ monotonically decreases with increasing $m^*$, which implies that, in the intrinsic limit, a smaller effective mass is better, as captured in Figure 3(a-b). Yet, through the Mott criterion, a larger effective mass tends to raise $N_D$, which dictates when the bulk state becomes metallic as a result of accumulation of charged defects. Since $N_D$ scales as $m^{*3}$, this dominates the overall dependence on $m^*$ as $P_{DS-BS} \propto 1/(1+\gamma m^{*3/2}) \approx 1 - \gamma m^{*3/2}$, in the limit of small $m^*$. Therefore, larger $m^*$ leads to topological materials that are more robust to defects and makes it easier to access the intrinsic limit, yet the finite temperature contribution of the bulk state will be larger. Effective mass strongly depends on the elements and bonding nature in the crystal through the band overlap. Typically, *d*- and *f*-orbitals are the most spatially localized, which gives rise to larger effective masses in comparison to the *s*- and *p*-orbitals that tend to have lower effective masses. Further, the heavier the elements are, the more delocalized the orbitals, which tends to give rise to smaller $m^*$. For comparison, several semiconductors are shown on the right-hand side of Figure 3(b). Since most topological materials typically involve *p*-block elements with strong covalency, the effective masses tend to be small, of order $0.2m_e$, which reduces thermally induced carriers and significantly reduces $N_D$. As shown in Figure 3(b), InSb has a low electron mass, $0.015m_e$, yet the hole mass is significantly larger at around $0.4m_e$. Since Equation (4) involves integrals over the conduction and valence bands, it may be a challenge to find materials with appreciably reduced electron and hole masses simultaneously; though, improvement can be made by pinning the Fermi level near the band with the lower effective mass. Alternatively, to access the intrinsic limit, searching for new TIs with 3d elements may produce larger $m^*$, which gives a net larger $N_D$. Yet, as discussed above, the bonds there tend to be more ionic, which both compromises band inversion while raising the ionic contribution to the dielectric constant. The former point has been well explored in the context of magnetic doping of tetradymites[10,38,39]. Altogether, this highlights that new materials are needed, but a careful balance among many parameters is required to reach the intrinsic limit and boost the maximum temperature for transport dominated by the topological states.

Concerning QAHIs, the results suggest observing quantized transport near room temperature will be very challenging, though certainly not impossible. The current materials paradigm for realizing a QAHI is (1) start with a 3D TI, (2) reduce the thickness, and (3) either interface it with a magnetic insulator or dope it to be magnetic[10]. Raising the temperature of a QAHI will require dramatic improvements to both the bulk material as well as the magnetic phase. In particular, the bulk band gap may need to be >0.5 eV with a magnetic gap >0.3 eV. This is a direct call for the continued development of dramatically different material paradigms[40–44]. Despite the challenge for pushing to room temperature, near liquid nitrogen temperatures may be within reach of current systems. This would be significantly improved over the current materials with maximum temperatures of 2-6 K[10,40,45]. Since the band gaps of the $(Bi_{1-x}Sb_x)_2Te_3$ and newly discovered $MnBi_2Te_4$ are of order of 0.15 meV, the 3D bulk should start to conduct near 75 K, which leaves the question of whether the magnetic gap can be increased, through either doping or heterostructuring. Given the extensive amount of work done, magnetic doping to open such a large gap is, again, unlikely, but interfacing such TIs with large-transition-temperature-ferromagnets should be a fruitful route to boosting the temperature up to this limit[32,33].

To conclude, the realization of practical topological electronic devices requires transport properties which are dominated by the topological states with minimal parallel conduction through the deleterious bulk. I have established simple criteria for a range of topological materials and discussed their dependence on chemical trends. At zero temperature, the conductivity of a 3D bulk state is dictated by the Mott criterion. Whereas in the intrinsic limit, where the Fermi level is at the center of the band gap, the limiting factor is the density of thermally induced carriers. For typical band gaps and effective masses this gives rise to significant bulk conduction at room temperature, which highlights that new material paradigms are needed. Specifically, operation at room temperature of a 3D TI can be achieved with a bulk band gap of around 0.5 eV, nearly twice as large as current materials. Excitingly, the combination of these parameters are calculatable with first principles methods, and, therefore, can be used to effectively parse databases of known and theorized topological materials[46]. This should enable narrowing down possible candidate materials for synthesis and characterization. However, current databases focusing on topological materials do not include these parameters. This highlights a critical gap in this field. Instituting this will enable rapidly identifying new candidate materials for initial synthesis and characterization, which will ultimately be incorporated into a practical topological device that could conceivably be used at room temperature for new technologies.

## Acknowledgments

This work was supported by the U.S. Department of Energy (DOE), Office of Science, Basic Energy Sciences, Materials Sciences and Engineering Division. I wish to thank Panchapakesan Ganesh, Jason Lapano, Joon Sue Lee, Seongshik Oh, Brian Sales, Brian Skinner, T. Zac Ward and Jie Zhang for insightful discussions, and T. Zac Ward for encouragement to publish this work.

## Conflict of Interest

The author declares no competing financial interest.

**Figures**

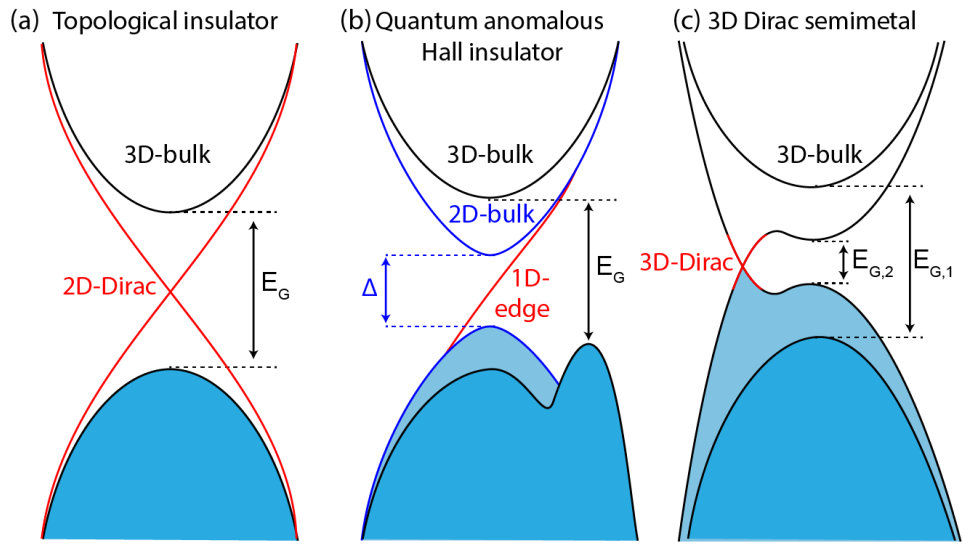

**Figure 1.** Schematic of topological band structures used for the calculations where $E_G$ is the bulk band gap, and $\Delta$ is the magnetic gap in the 2D-topological surface states. (a) Topological insulators after $Bi_2Se_3$[1], (b) Quantum anomalous Hall insulator after magnetically-doped $(Bi,Sb)_2Te_3$[10,47], and (c) 3D Dirac semimetal after $Cd_3As_2$[48]. Black curves: 3D bulk states, Blue curves: 2D bulk states, Red curves: topological Dirac states.

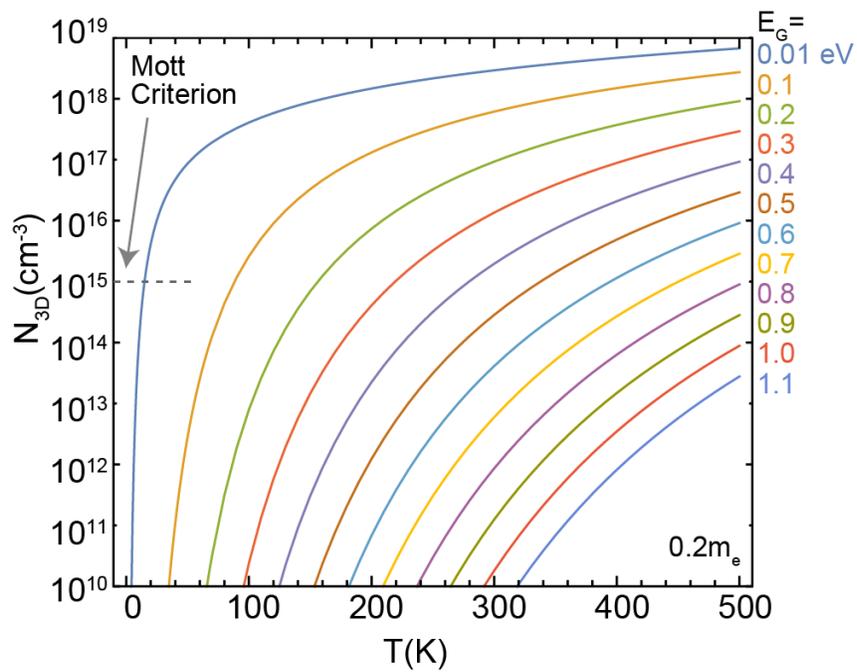

**Figure 2**. (a) Temperature dependence of the carrier density for various band gaps with $m^* = 0.2m_e$ for intrinsic materials.

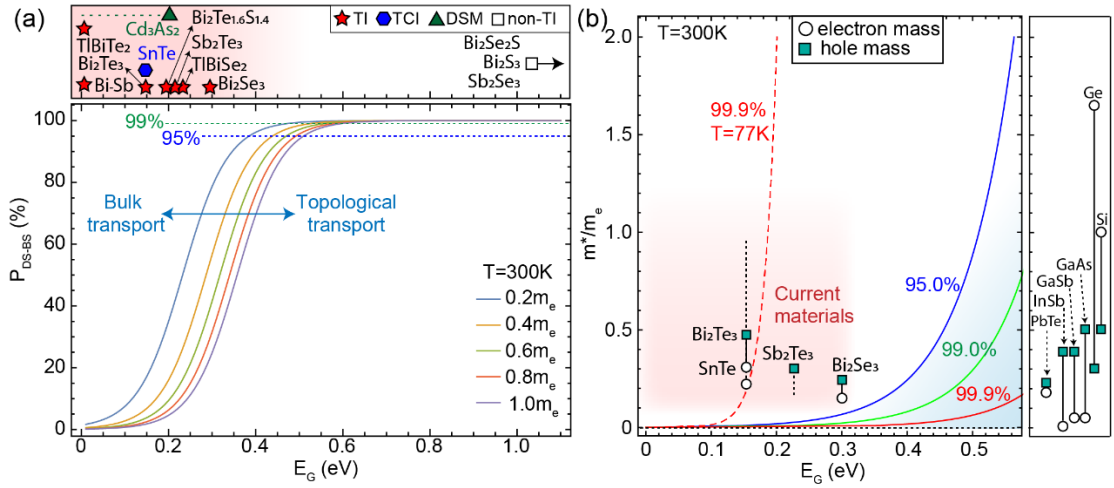

**Figure 3**. (a) Percent of conduction that arises from the Dirac state, $P_{DS-BS}$, plotted versus band gap for various effective masses, as indicated. This is calculated for a 2D Dirac state of 20 nm or a 3D Dirac state. For comparison, the horizontal dashed lines indicate where $P_{DS-BS}$=95% and 99%, as well as for select topological materials in the top panel. (b) Set of band gap versus effective mass where $P_{DS-BS}$=95% (blue), 99% (green) and 99.9% (red). The shaded area in the lower right indicates where the topological states dominate transport. The dashed red curve is $P_{DS-BS}$=99.9% at $T$=77 K. Effective masses for various materials are included for comparison (for materials with both heavy and light bands the heavy band is shown). Data for select materials is from Bi-Sb[49], TlBiTe$_2$/TlBiSe$_2$[50,51], Bi$_2$Se$_3$/Bi$_2$Te$_3$/Sb$_2$Te$_3$[16], SnTe [52], Cd$_3$As$_2$[48], Bi$_2$Te$_{1.6}$Se$_{1.4}$[53], Bi$_2$Se$_2$S[54], Bi$_2$S$_3$[55], Sb$_2$Se$_3$[56], Si/Ge/GaAs/GaSb/InSb/PbTe[27].

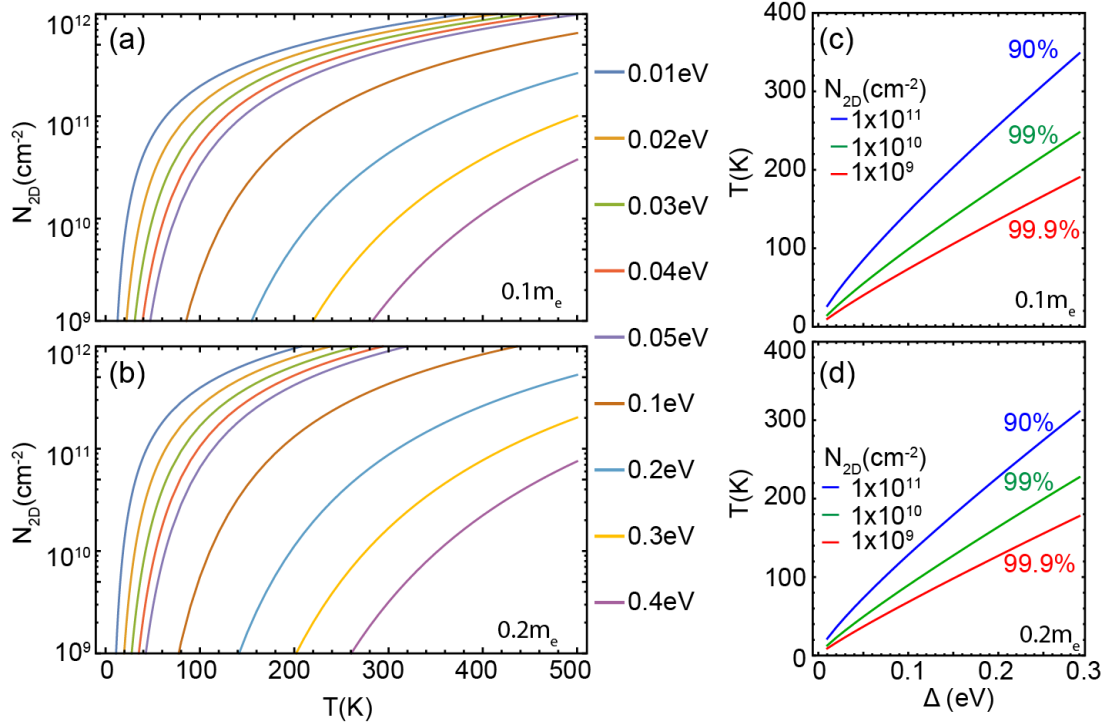

**Figure 4.** (a-b) Temperature dependence of the 2D carrier concentration for various band gaps with $m^* = 0.1 m_e$ (a) and $m^* = 0.2 m_e$ (b). (c-d) Magnetic gap versus temperature where the curves in (a-b) intersect the carrier concentrations of $1\times10^{10}$ cm$^{-2}$ (blue), $1\times10^{9}$ cm$^{-2}$ (green), $1\times10^{8}$ cm$^{-2}$ (red), which, as indicated, correspond to percent transport of 90%, 99% and 99.9%, respectively, through the topological state.

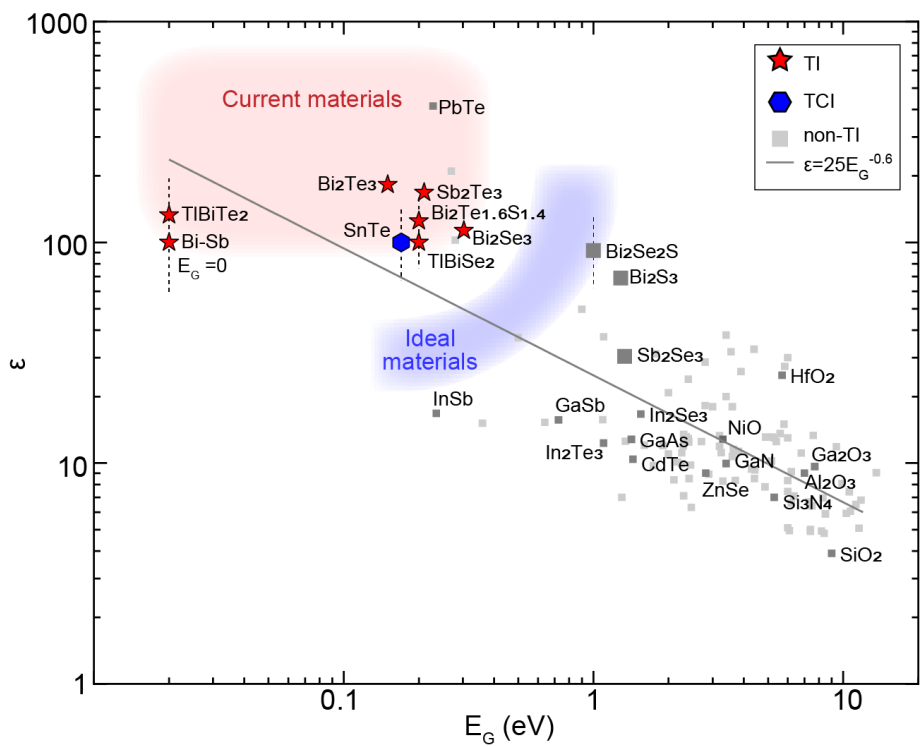

**Figure 5.** Compilation of bandgap versus zero frequency dielectric constant for topological materials as well as non-TIs. A vertical dashed line signifies $\epsilon$ is not reported. Data is from Bi-Sb[49], TlBiTe$_2$/TlBiSe$_2$[50,51], Bi$_2$Se$_3$/Bi$_2$Te$_3$/Sb$_2$Te$_3$[16], SnTe[52], Bi$_2$Te$_{1.6}$Se$_{1.4}$[53], Bi$_2$Se$_2$S[54], Bi$_2$S$_3$[55,57], Sb$_2$Se$_3$[37,56], and non-TIs (small squares) and fit (solid line)[37].